\documentclass[a4paper]{article}

\usepackage[margin=1in]{geometry} % full-width

\usepackage{amsmath}
\usepackage{amsthm}
\usepackage{amssymb}

\usepackage[utf8]{inputenc}
\usepackage{hyperref}
\hypersetup{
	unicode,
	pdfauthor={},
	pdftitle={Programmable Integrated Photonics for Topological Hamiltonians},
	pdfsubject={},
	pdfkeywords={},
	pdfproducer={LaTeX},
	pdfcreator={pdflatex}
}

\theoremstyle{plain}

\theoremstyle{definition}

\usepackage{graphicx, color, xcolor}
\graphicspath{{fig/}}

\usepackage{algorithm, algpseudocode} % use algorithm and algorithmicx for typesetting algorithms
\usepackage{mathrsfs} % for \mathscr command

\usepackage{lipsum}
\usepackage{cite}

\usepackage[normalem]{ulem}
\usepackage[figurename=Fig.]{caption}

\newcommand{\add}{\textcolor{black}}
\newcommand{\remove}[1]{}

\title{A programmable photonic memory}
\author{Farshid Ashtiani$^{1,*}$}

\date{
	$^1$\textit{Nokia Bell Labs, 600 Mountain Ave, Murray Hill, NJ 07974, USA} \\%
    $^*$farshid.ashtiani@nokia-bell-labs.com\\[2ex]%
}

\begin{document}
\maketitle

\begin{abstract}
The significant advancements in integrated photonics have enabled high-speed and energy efficient systems for various applications from data communications and high-performance computing, to medical diagnosis, sensing and ranging. However, data storage in these systems has been dominated by electronic memories which necessitates signal conversion between optical and electrical as well as analog and digital domains, and data movement between processor and memory that reduce the speed and energy efficiency. To date, a scalable optical memory with optical control has remained an open problem. Here we report an integrated photonic set-reset latch as a fundamental optical static memory unit based on universal optical logic gates. While the proposed memory is compatible with different photonic platforms, its functionality is demonstrated on a programmable silicon photonic chip as a proof of concept. Optical set, reset, and complementary outputs, scalability to a large number of memory units via the independent latch supply light, and compatibility with different photonic platforms enable more efficient and lower latency optical processing systems. 
\end{abstract}

	%\tableofcontents

%%%%%%%%%%%%%%%%%%%%%%%%%%%%% INTRO %%%%%%%%%%%%%%%%%%%%%%%%%%%%%%%

	\section{Introduction}

\begin{figure}[ht!]
\centering\includegraphics[width=\linewidth]{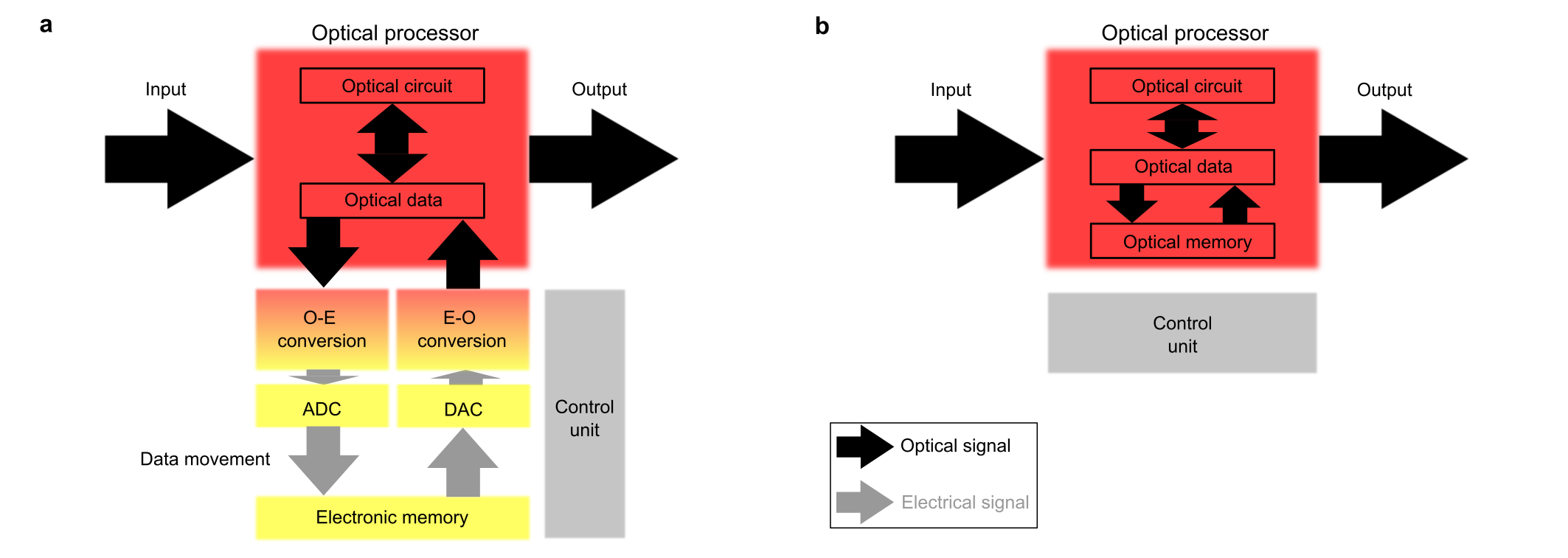}
\caption{\textbf{Comparing electronic and optical memory interfaces in an optical processor.} \textbf{a} An optical processor with electronic memory. Optical data to be stored in the memory should be photo-detected, digitized, and transferred to the electronic memory. Similarly, to retrieve data, it should be transferred and converted to analog signals that modulate an optical carrier to up-convert the data to optical domain to be processed by the system. \textbf{b} Optical processor with embedded optical memory. E-O and O-E conversions, ADC and DAC, and electronic data movement can be eliminated, resulting in lower latency, higher energy efficiency, smaller the system size, and less packaging complexities.}
\label{fig1}
\end{figure}

To support a sustainable growth of information technology, there have been great efforts towards faster and more energy efficient computation, communication, and storage systems, the three main pillars of information technology. While electronics has been the main driving force in all three aspects, a sustained increase in speed and energy efficiency has become challenging. The inherent loss mechanisms in electrical transmission lines that make data movement costly, as well as the limited clock frequency in digital processors that ultimately determines the computation speed limits, are among the major reasons. Hence, optics has gained even more attention as a promising candidate to address some of these challenges. Large available optical bandwidth, low loss signal transmission, various multiplexing schemes (such as time, wavelength, mode, polarization), and the recent advancements in chip integration and packaging have enabled higher throughput and energy efficient systems. Nevertheless, these systems mainly target the first two pillars of communication \cite{opt_comm1, opt_comm2,opt_comm3} and computation and processing systems \cite{opt_proc1,opt_proc2,opt_proc3,opt_proc4}. Storage systems, as the third main pillar, have been predominantly realized using digital electronics \cite{opt_proc2}. The bosonic nature of photons makes the realization of electronic-like (that is, electric charge-based) optical memories impractical. Therefore, to store optical data on an electronic memory, light should be photo-detected (optical to electrical, O-E, conversion), digitized (analog to digital conversion, ADC), and transferred to the memory (data movement). Similarly, to read and process the memory data in the optical processor, data transfer from memory, digital to analog conversion (DAC), and electrical to optical (E-O) conversion through optical modulation are required, as shown in Fig. \ref{fig1}a. All of these steps of O-E/E-O conversion, ADC/DAC, and data movement reduce the energy efficiency, increase processing latency, and complicate system integration and packaging of the overall optical processor \cite{opt_proc2}. Therefore, an optical memory where optical data can be directly stored and retrieved with no conversion can benefit the system performance. Figure \ref{fig1} schematically summarizes the above discussion and compares two optical processors with and without embedded optical memory.

Various volatile and non-volatile optical memories have been proposed to date \cite{mem_survey,udisk1,uring1, uring2,SOA_MZI1,SOA_MZI2,SOA_XPM1,PHC1,PHC2,PHC3,loop1,loop2,dope1,dope2,PCM1,PCM2,PCM3}. Static volatile memories, where data refresh is not required, have been proposed mainly based on optical bistability. Methods include using coupled micro-disk \cite{udisk1} and micro-ring lasers \cite{uring1, uring2}, semiconductor optical amplifiers in a Mach-Zehnder interferometer (MZI) \cite{SOA_MZI1,SOA_MZI2} and through cross-phase modulation \cite{SOA_XPM1}, and photonic crystals with buried cavities \cite{PHC1,PHC2,PHC3}. Dynamic optical memories with required data refresh have been proposed using recirculating fiber loops \cite{loop1,loop2} as well as doped fibers \cite{dope1,dope2}. In addition, non-volatile optical memories have been primarily implemented using various phase change materials \cite{PCM1,PCM2,PCM3}. Despite impressive demonstrations, most of prior art either rely on bench-top setups \cite{uring2,SOA_MZI1,SOA_MZI2,SOA_XPM1,loop1,loop2,dope1,dope2}, or require certain material systems \cite{udisk1,uring1,PHC1,PHC2,PHC3,PCM1,PCM2,PCM3} that both make large scale integration more costly and challenging. For an optical memory to be a practical alternative to an electronic counterpart, large scale, high-yield, and low-cost fabrication with small physical size are required. In the past few years, silicon photonics has become the main platform offering such features \cite{siph1,siph2,siph3}. Therefore, a silicon photonic compatible solution enables large-scale integration as well as co-integration and co-design with the existing silicon photonic communication and computing systems.  

Here we demonstrate a novel and scalable integrated photonic set-reset latch (SR latch) as an optical static memory unit, based on universal optical logic gates (UOLGs), namely NOR and NAND. The UOLGs are realized based on the nonlinear electro-optic response of integrated silicon micro-ring modulators (MRMs) and can be scaled to several gates. By combining the UOLGs, more complex optical logic operations can be achieved and here, we demonstrate an optical SR latch with optical set, reset, and complementary output signals. Our optical SR latch can be implemented on any integrated photonic platform that offers modulators and detectors. Programmable photonic platforms based on a mesh of MZIs enable the realization and real-time reconfiguration of a wide variety of photonic circuits \cite{prog_ph1}. As a proof of concept, here we use iPronics SmartLight processor \cite{ipron1,ipron2,ipron3}, a commercially available programmable silicon photonic chip, to demonstrate accurate operation of the proposed UOLGs and the SR latch memory unit. Compared to the previous optical memories, our method can be co-integrated with the existing silicon photonic processing systems within a small footprint with no further post processing and no bulky bench-top setup required. Hence, this is an important step towards scalable optical memories for a variety of photonic systems.

%%%%%%%%%%%%%%%%%%%%%%%%%%%%% UOLG %%%%%%%%%%%%%%%%%%%%%%%%%%%%%%%

\section{Photonic memory unit based on universal logic gates}

\begin{figure}[t!]
\centering\includegraphics[width=\linewidth]{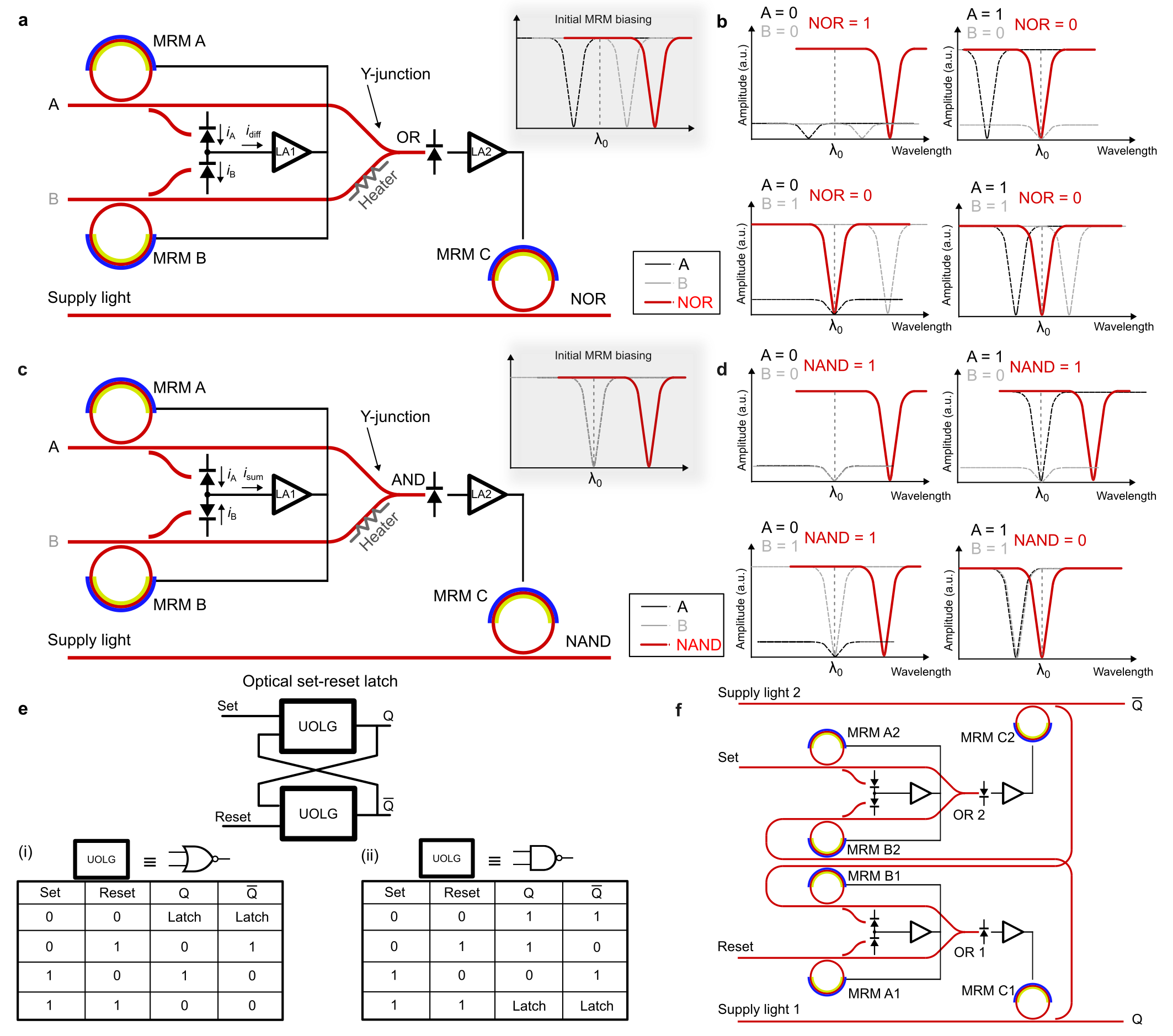}
\caption{\textbf{Universal optical logic gates.} \textbf{a} Photonic NOR circuit consisting of two MRMs to generate the maximum of the two inputs (OR operation) followed by MRM C to perform NOT operation and \textbf{b} Transmission spectra for different logic input combinations. \textbf{c} Photonic NAND circuit consisting of two MRMs to generate the minimum of the two inputs (AND operation) followed by MRM C to perform NOT operation and \textbf{d} Transmission spectra for different logic input combinations. \textbf{e} Schematic of an optical SR latch and the corresponding truth tables using (i) two NOR and (ii) two NAND gates. \textbf{f} Photonic SR latch circuit using two cross-coupled NOR gates.}
\label{fig2}
\end{figure}

 Figures \ref{fig2}a and \ref{fig2}c show the proposed architectures for optical NOR and NAND gates using the nonlinear electro-optic response of MRM. Previously, we proposed a similar approach to implement photonic max-pooling function for optical neural networks \cite{maxpool}. In Fig. \ref{fig2}a, two optical inputs A and B at wavelength $\lambda_{0}$ are coupled to MRM A and MRM B, respectively. The MRMs are initially biased such that their corresponding resonance wavelengths are symmetrically positioned around $\lambda_{0}$ (the shaded panel in Fig. \ref{fig2}a). This is performed by first aligning the resonance wavelengths of both MRMs with $\lambda_{0}$ and then thermally tuning them to achieve the desired biasing condition. After each MRM, a small fraction of the optical signal is tapped off and coupled to a photodetector (PD). The two photocurrents are subtracted and the current difference $i_{diff} = i_{A} - i_{B}$ is amplified using the limiting amplifier 1 (LA1), driving MRM A and MRM B. The rest of the optical signals are combined using a Y-junction. As shown in Fig. \ref{fig2}b, the resonance wavelengths of MRM A and MRM B shift depending on the logic state of inputs A and B. The LA1 gain is designed such that for $i_{diff}>0$ (that is, A = high and B = low), the resonance wavelength of MRM B is aligned with $\lambda_{0}$ while MRM A is still off-resonance, resulting in further attenuation of B. Similarly, $i_{diff}<0$ (that is, A = low and B = high) results in further attenuation of A. Moreover, when both inputs are high (low), the Y-junction output is also high (low). This can be ensured by properly adjusting the heater placed in one of the two signal paths. Effectively, the optical output after the Y-junction corresponds to the maximum of A and B, which is equivalent to an optical logic OR operation. To realize an optical NOR function as a UOLG, a logic NOT operation is required. This is achieved using MRM C with an independent supply light at $\lambda_{0}$. MRM C resonance wavelength is first aligned with $\lambda_{0}$ and then thermally tuned to maximize the MRM transmission. The OR output after the Y-junction is photo-detected and amplified using LA2. The LA2 gain is designed such that a high (logic `1`) OR output results in aligning the resonance of MRM C with $\lambda_{0}$ (attenuating the supply light) and a low (logic `0`) OR output does not shift the resonance. Therefore, MRM C inverts the OR output and, together, the OR and the NOT functions form an optical NOR gate which is a UOLG that can be used to implement an SR latch. 

Similarly, an optical NAND gate can be made using an AND followed by a NOT gate as shown in Fig. \ref{fig2}c. This architecture is almost identical to the one in Fig. \ref{fig2}a except for two differences. First, MRMs A and B are initially biased such that both resonance wavelengths are aligned with $\lambda_{0}$ (the shaded panel in Fig. \ref{fig2}c). Furthermore, instead of generating a current difference, the sum of photocurrents $i_{sum} = i_{A} + i_{B}$ is generated. In this case, if both inputs are high, $i_{sum}$ is maximized. The LA1 gain is designed such that for maximum value of $i_{sum}$, MRMs A and B are shifted to become off-resonance. This results in a high optical power at the Y-junction output. For any other input scenario, $i_{sum}$ is not high enough to shift the MRMs. Therefore, the Y-junction output follows the minimum of the inputs which is equivalent to a logic AND operation. A similar NOT gate as in Fig. \ref{fig2}a inverts the AND output, generating the optical NAND output. Note that due to the use of the independent supply light the output optical power can be maintained at a certain level within each gate and the optical loss does not propagate when multiple gates are cascaded. Hence, the proposed NOR/NAND gate can be scaled to large number gates for more complex logic functions.  

By using two NOR/NAND gates, an SR latch memory unit can be formed. Figure \ref{fig2}e shows the top level schematic of a photonic SR latch as an optical memory unit, as well as the truth tables in the cases of using (i) two NOR and (ii) two NAND gates. Here, two cross-coupled UOLGs, each with two optical inputs and one optical output, are used to set, reset, or latch (hold) the state of the memory. Figure \ref{fig2}f shows the architecture of the proposed optical SR latch based on two NOR gates (from Fig. \ref{fig2}a), where set, reset, and the two complementary outputs \textit{Q} and $\overline{Q}$ are all in the optical domain. In this work, we use two NOR gates to demonstrate the latch functionality. It is worth mentioning that in the proposed architecture, the memory response time is a function of the electro-optic and opto-electronic bandwidths of the MRM and the PD, respectively. Considering that in most commercially available silicon photonic processes modulators and detectors with bandwidths of above 30 GHz are available, picosecond-scale memory response time can be achieved.

%%%%%%%%%%%%%%%%%%%%%%%%%%%%% Latch %%%%%%%%%%%%%%%%%%%%%%%%%%%%%%%

\section{Optical memory on a programmable photonic platform}

\begin{figure}[t]
\centering\includegraphics[width=\linewidth]{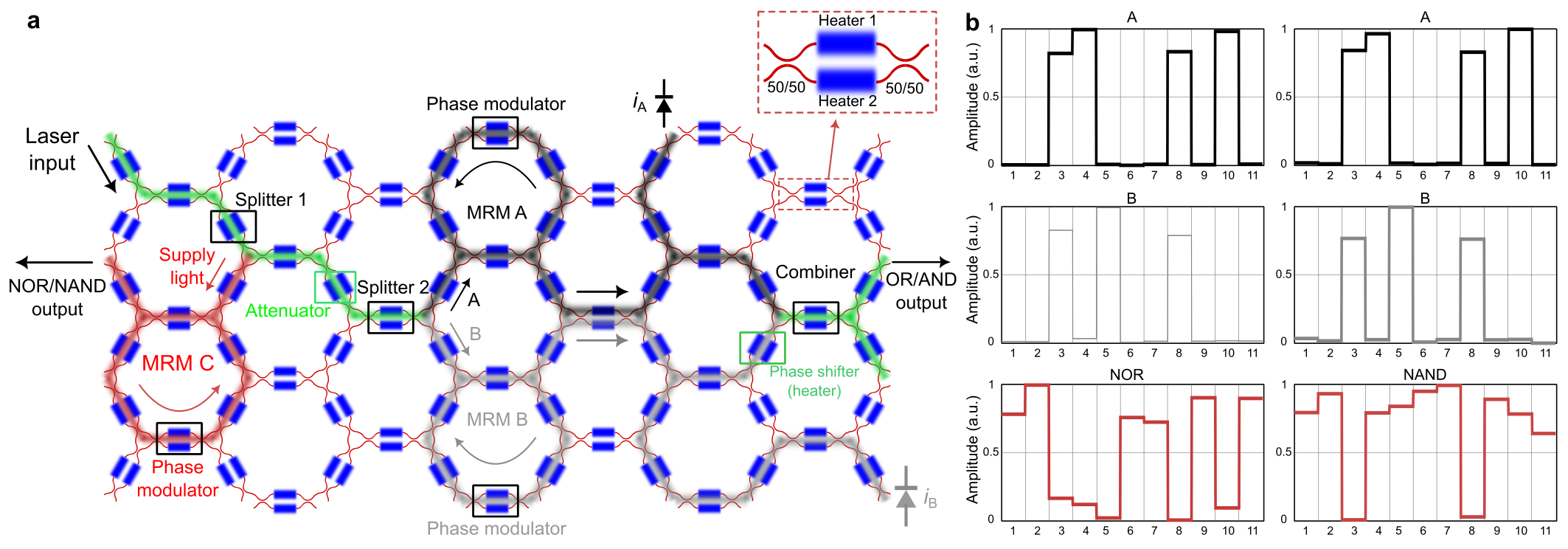}
\caption{\textbf{Implementation of NOR/NAND gates on the hardware mesh.} \textbf{a} Schematic of the hardware mesh made with 2$\times$2 MZIs. The signal path for signals A, B, and supply light are marked with black, grey, and red lines, respectively. Green lines mark the paths with multiple signals. \textbf{b} Experimental results for NOR and NAND gates. Arbitrary amplitude variations are added to the input signals to ensure robustness. In each graph, vertical axis shows the normalized amplitude and horizontal axis shows the input sequence.}
\label{fig3}
\end{figure}

As a proof of concept, the proposed photonic NOR gate, NAND gate, and SR latch memory unit are implemented on iPronics SmartLight programmable MZI mesh. Figure \ref{fig3}a shows the implementation of optical NOR/NAND gates on the programmable photonic chip for experimental verification. Each 2$\times$2 MZI is formed by two 50/50 directional couplers and two thermal phase shifters, one on each arm of the MZI, that can be programmed to any split ratio between 0 and 1 (zoomed-in picture in Fig. \ref{fig3}a). All MZIs within the mesh can be programmed using a control software and the peripheral ports can be used to photo-detect the desired signals. More details about the hardware mesh can be found in Methods section. Note that since NOR and NAND gates are only different in the feedback signal ($i_{diff}$ vs. $i_{sum}$) and initial MRM biasing conditions that are both defined in the control software, the physical circuits on the hardware are identical. Here, the system has one optical input and it is used to generate A, B, and the supply light in Fig. \ref{fig2}a and \ref{fig2}c (see Supplementary Note 1 for the details of this implementation). The laser source is coupled to the top-left port of the mesh. The input light is first split into two signals using splitter 1 to generate the supply light for the NOT function. Splitter 2 is then used to generate the logic inputs A and B where by programming the split ratio, various logic input combinations can be generated. The scenario in which both inputs are low can be mimicked by attenuating the signal before splitter 2. The attenuator can be formed by changing the split ratio in an MZI switch and taking the output from one port. Three MRMs are formed by programming six MZI arms, with the desired coupling ratio for each ring resonator. One MZI arm in each MRM is used to modulate the optical phase within the ring and to apply the feedback control signal (that is, the output of the LAs in Fig. \ref{fig2}). After MRMs A and B, 5\% of the signals are tapped off, routed towards the peripheral ports, and photo-detected to generate $i_{A}$ and $i_{B}$. LA1 and LA2 that generate the feedback signals are implemented in the control software. The two optical paths are then combined using the combiner to generate the OR/AND output. As explained in the previous section, the OR/AND output is photo-detected and amplified to drive the phase modulator of MRM C and generate the NOR/NAND output. A simplified schematic of this circuit is shown in Fig. S1 of Supplementary Note 1.

Figure \ref{fig3}b shows the experimental results for NOR and NAND UOLGs. A sequence of different combinations of logic inputs A and B is generated by properly programming the relevant splitters. To ensure the robustness of UOLGs, arbitrary variations are added to optical logic levels `0` and `1` by adding random variation to split ratios. It can be seen that despite the variations, both gates generate the correct logic output for all input scenarios. Details of the measurement setup are presented in Methods section.

\begin{figure}[t]
\centering\includegraphics[width=\linewidth]{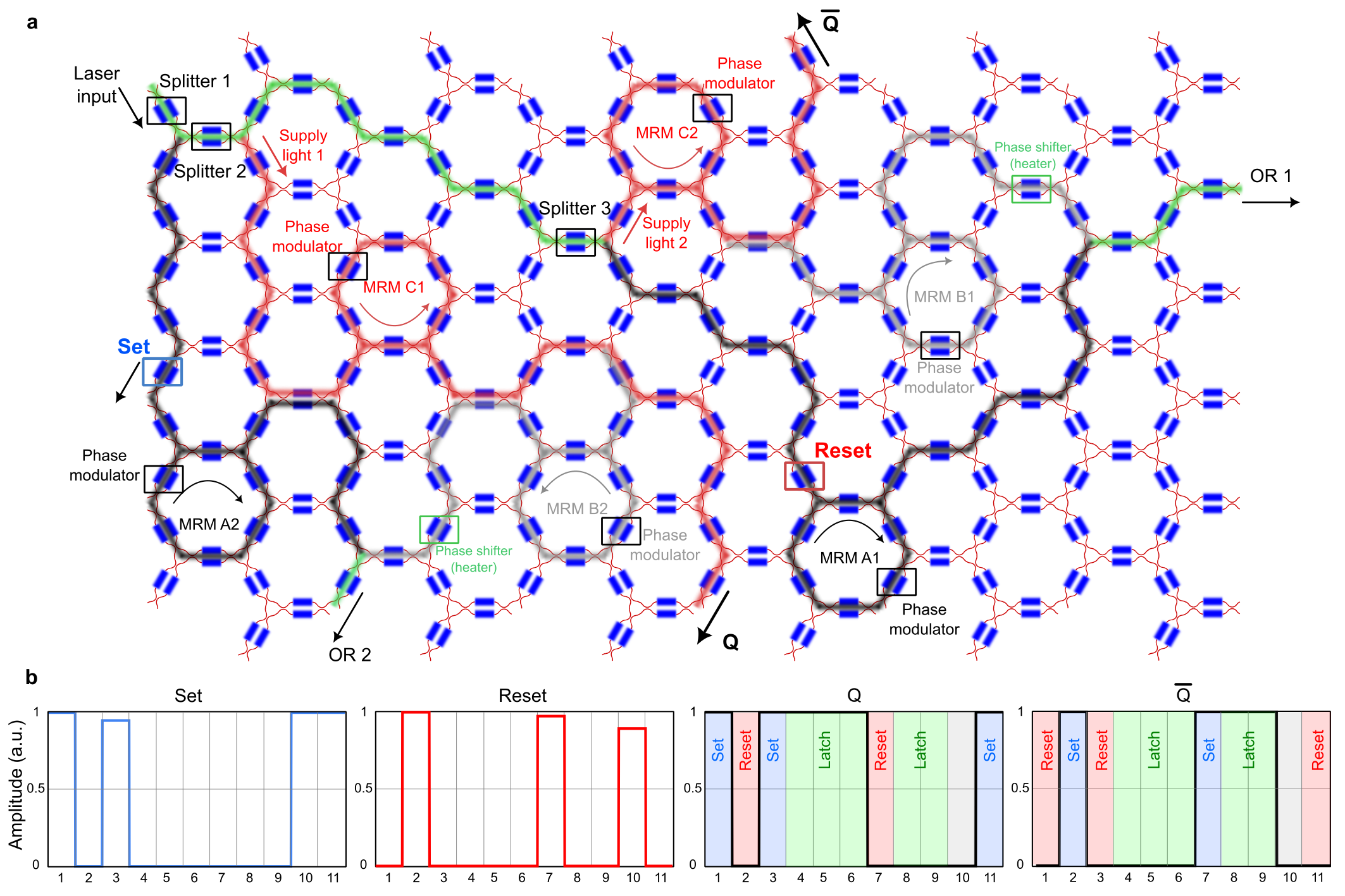}
\caption{\textbf{Optical SR latch on programmable MZI mesh.} \textbf{a} Schematic of the simulator mesh made with 2$\times$2 MZIs. The mesh size allows for having two coupled UOLGs. For each gate, the signal path for signals A, B, and supply light are marked with black, grey, and red lines. Set and reset attenuators are shown in blue and red, respectively. Green lines mark the paths with multiple signals. \textbf{b} Simulation results of the SR latch circuit. Different combinations of set (blue) and reset (red) signals are generated and the corresponding complimentary outputs $Q$ and $\overline{Q}$ are plotted.}
\label{fig4}
\end{figure}

Two UOLGs can then be cross-coupled to form an optical SR latch memory. However, the size and shape of the currently available hardware platform (Fig. \ref{fig3}a) is such that only one UOLG fits the mesh. Nevertheless, the SmartLight processor also offers a realistic simulator with larger mesh size \cite{ipron_sim,topo_paper} that is suitable for implementing the SR latch. The simulator provides a very close approximation of the actual hardware \cite{ipron_sim}. Note that the mesh size in the simulator is completely within the current silicon photonics fabrication capabilities. More information on the SmartLight processor hardware and simulator can be found in Methods section.

While either photonic NOR or NAND gates can be used in the latch architecture, here we use two photonic NOR gates to form the optical SR latch as previously shown in Fig. \ref{fig2}f. Figure \ref{fig4}a shows the proposed latch formed by two coupled photonic NOR gates on the simulator. Similar to Fig. \ref{fig3}a, the SR latch is implemented using one optical input (see Supplementary Note 2 for the details of this implementation). In Fig. \ref{fig4}a, splitters 1 to 3 generate supply lights 1 and 2 as well as the set and reset signals. Set and reset states are defined after passing through variable attenuators (marked with blue and red boxes, respectively). MRMs C1 and C2 perform the NOT function and MRMs A1, B1, A2 and B2 generate the OR1 and OR2 outputs. Similar to Fig \ref{fig2}f, splitters after MRM C1 and C2 are used to monitor the latch outputs $Q$ and $\overline{Q}$. The biasing conditions of the MRMs are similar to those of Fig. \ref{fig2}a. The required feedback signals are generated using the control software and applied to the relevant MRMs. See Fig. S2 of Supplementary Note 2 for a simplified schematic and the details of cross-coupling two NOR gates in the latch circuit of Fig. \ref{fig4}a.

To demonstrate the performance of the proposed photonic SR latch, different set and reset combinations are generated using the corresponding attenuators in Fig. \ref{fig4}a. Figure \ref{fig4}b shows the latch simulation results. Eleven different input combinations are sequentially generated and the corresponding complementary latch outputs are shown. Set, reset, and latch states are shaded in blue, red, and green, respectively. The grey region corresponds to the case where both set and reset are defined as logic `1` which results in logic `0` for both $Q$ and $\overline{Q}$. It can be seen that the proposed programmable photonic memory unit performs all functions of set, reset, and latch/hold accurately. Similar to UOLG experiments, amplitude variations are added to the set and reset signals and despite these variations, the latch outputs maintain the accurate logic levels. In addition to the effect of using independent supply lights as explained before, due to the reinforcing behavior (positive feedback mechanism) of the cross-coupled architecture of the SR latch, the low and high logic levels of $Q$/$\overline{Q}$ are very close to 0 and 1, respectively, compared to the results in Fig. \ref{fig3}b. The details of the realistic simulator and the simulation conditions are provided in Methods section.

%%%%%%%%%%%%%%%%%%%%%%%%%%%%% Summary %%%%%%%%%%%%%%%%%%%%%%%%%%%%%%%

\section{Discussion and conclusion}

The scalibility of the architectures presented in this work, compared to the previous demonstrations, originates from both fabrication and circuit design aspects. Compatibility with silicon photonic platforms without the need for any additional material system, post-processing, and bulky bench-top setups, enables low-cost and large-scale fabrication within a small footprint. Moreover, fabrication processes that offer co-integration of electronic and photonic components \cite{45clo} enable further size reduction, scalability, and higher efficiency by simplifying the packaging requirements. This can be achieved by co-designing and co-integrating the electronic with the photonic circuits to generate the feedback signals on a single chip which minimizes the need for communicating with off-chip modules. This also reduces the parasitic elements which improves the bandwidth and latency. While our proof of concept demonstration uses a programmable chip with thermal phase shifters, using high-speed electro-optic modulators and detectors available in various integrated photonic platforms \cite{fast_prog} enables memory response times of tens of picosecond \cite{opt_proc3}. In addition to fabrication process related aspect, the use of independent per memory unit supply light allows for larger scale systems by maintaining a certain optical power range within each memory unit, independent of the others. 

In summary, we proposed and demonstrated a novel scalable photonic SR latch as an optical memory unit. The latch operates based on two cross-coupled UOLGs that use the nonlinear electro-optic response of MRMs. Photonic NOR and NAND gates are experimentally demonstrated on iPronics SmartLight programmable silicon photonic platform and all latch functions (set, reset, hold) are demonstrated using iPronics realistic simulator with larger mesh size than the currently available hardware. Optical data storage is one of the major challenges that optical processing systems face, and this work is an important step towards an scalable solution that can be co-designed and integrated with the existing and future integrated photonic processors. 

%%%%%%%%%%%%%%%%%%%%%%%%%%%%% Methods %%%%%%%%%%%%%%%%%%%%%%%%%%%%%%%

\section{Methods}
\textbf{Hardware and software calibration process}\\ Proper calibration of the MRM biasing conditions is essential to achieve optical logic functions and hence, the SR latch. Therefore, a calibration phase is required to initialize the mesh. First, the mesh is programmed using Python interface to create the desired photonic circuit architecture. This is performed by setting the split ratio of the MZI arms to route the optical signal in the desired path (Figs. \ref{fig3} and \ref{fig4}). Given the input laser wavelength $\lambda_{0}$, the resonance wavelengths of MRMs A and B in Fig. \ref{fig3} are first aligned with $\lambda_{0}$. The two MRMs are kept aligned with $\lambda_{0}$ for the NAND gate, and thermally tuned by a predefined amount in opposite directions for the NOR gate. A similar process is used for the NOT function (MRM C) to allow for high transmission when OR/AND signals are low and low transmission when OR/AND are high, to invert the OR/AND output. In addition, the heater shown in Figs. \ref{fig2}a and \ref{fig2}c should be properly adjusted to set the relative phase between the two signals paths in NOR/NAND gate. Note that the calibration is required both for the hardware and the simulator to initialize the rings. In the case of the hardware mesh, it is also required to compensate for any fabrication and environmental variations.  \\ \\
\textbf{Experimental setup}\\ The setup includes a lasers source, a polarization controller, and the iPronics SmartLight processor. In this case, the internal laser included in the processor box is used with an external polarization controller. The laser is tunable from 1549.9\,nm to 1550.2\,nm, and its wavelength is set to about 1550\,nm. Light is coupled with a fiber to chip coupling loss of about 2-4\,dB. The hexagonal mesh consists of 72 programmable 2$\times$2 splitters, each  811\,$\mu$m long with an average insertion loss of about 0.5\,dB. Each splitter is formed by a 2$\times$2 MZI with heaters as optical phase modulators. MZI splitters can be individually programmed using a Python interface which enables real-time control over the photonic circuit. The processor features off-chip PDs with a sensitivity of -70\,dBm that are used for feedback control signal generation as well as monitoring purposes. The photonic chip is temperature stabilized within $\pm$0.5$^\circ$C of room temperature.  \\ \\
\textbf{Realistic simulator mesh}\\ The hardware mesh size (72 MZI splitters) is large enough to fit one UOLG. Since the latch requires two UOLGs, a mesh with twice larger number of splitters is required. Therefore, we have used the realistic simulator provided with the hardware to implement the latch. In this case, 198 MZI splitters are available that are programmed similar to the ones on the chip. To have a realistic simulation, splitters are defined lossy similar to the hardware. The simulator mesh size and characteristics are within the technological capabilities of silicon photonics. Due to the similarity of simulator programming to that of the hardware, the implemented optical latch can be easily transferred to a hardware mesh with an adequate size.

\add{
\section{Data availability}
The data supporting findings of this study are available from the corresponding author upon reasonable request.
}

%%%%%%%%%%%%%%%%%%%%%%% References %%%%%%%%%%%%%%%%%%%%%%%%%

%%%%%%%%%% If using BibTeX:
%\bibliographystyle{opticajnl}
\def\url#1{}
\bibliographystyle{naturemag}
\bibliography{refs}
%\addbibresource{refs.bib}

\section{Acknowledgments}
The author would like to thank iPronics for technical support and Maryam Daniali for insightful discussions.

\section{Author contributions}
F.A. conceived the idea, conducted the experiments and simulations, analyzed the results, and wrote the paper.

\section{Competing interests}
The author declares no competing interests.

\section{Additional information}
\textbf{Correspondence} should and requests for materials should be addressed to Farshid Ashtiani.

\newpage

\section*{Supplementary Note 1: Universal optical logic gates circuit on the hardware mesh using one optical input}

As mentioned in the manuscript, one optical input is used to couple light to the hardware mesh. Hence, the same input is used to generate inputs A and B as well as the supply light. To illustrate this more clearly, Fig. \ref{Fig_S1} shows the schematic of the circuit in Fig. 3a of the main text. The signal and component names are the same as shown in the main text. Here, the input light is split into two parts, one as the supply light to MRM C for logic NOT operation, and one to generate the inputs A and B to the gate using splitter 2. For better control over optical levels and to implement the scenario where both inputs are low, an attenuator is placed before splitter 2. The OR/AND output is routed to the peripheral ports to be photo-detected and to drive MRM C to invert OR/AND and generate NOR/NAND output. 

\setcounter{figure}{0}
\renewcommand{\thefigure}{S\arabic{figure}}
\begin{figure}[ht!]
\centering\includegraphics[width=1\linewidth]{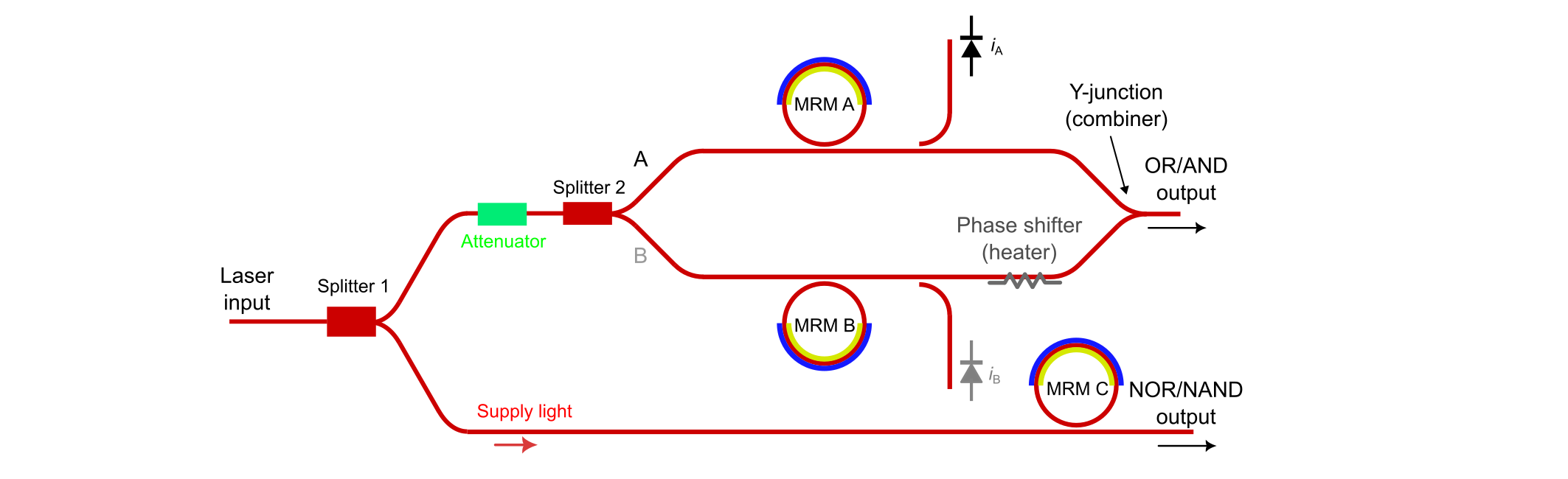}
\caption{Schematic of the NOR/NAND circuit implemented on the hardware mesh shown in Fig. 3a of the main text.}
\label{Fig_S1}
\end{figure}

\newpage
\section*{Supplementary Note 2: Photonic SR latch circuit on the simulator mesh using one optical input}

\begin{figure}[ht!]
\centering\includegraphics[width=1\linewidth]{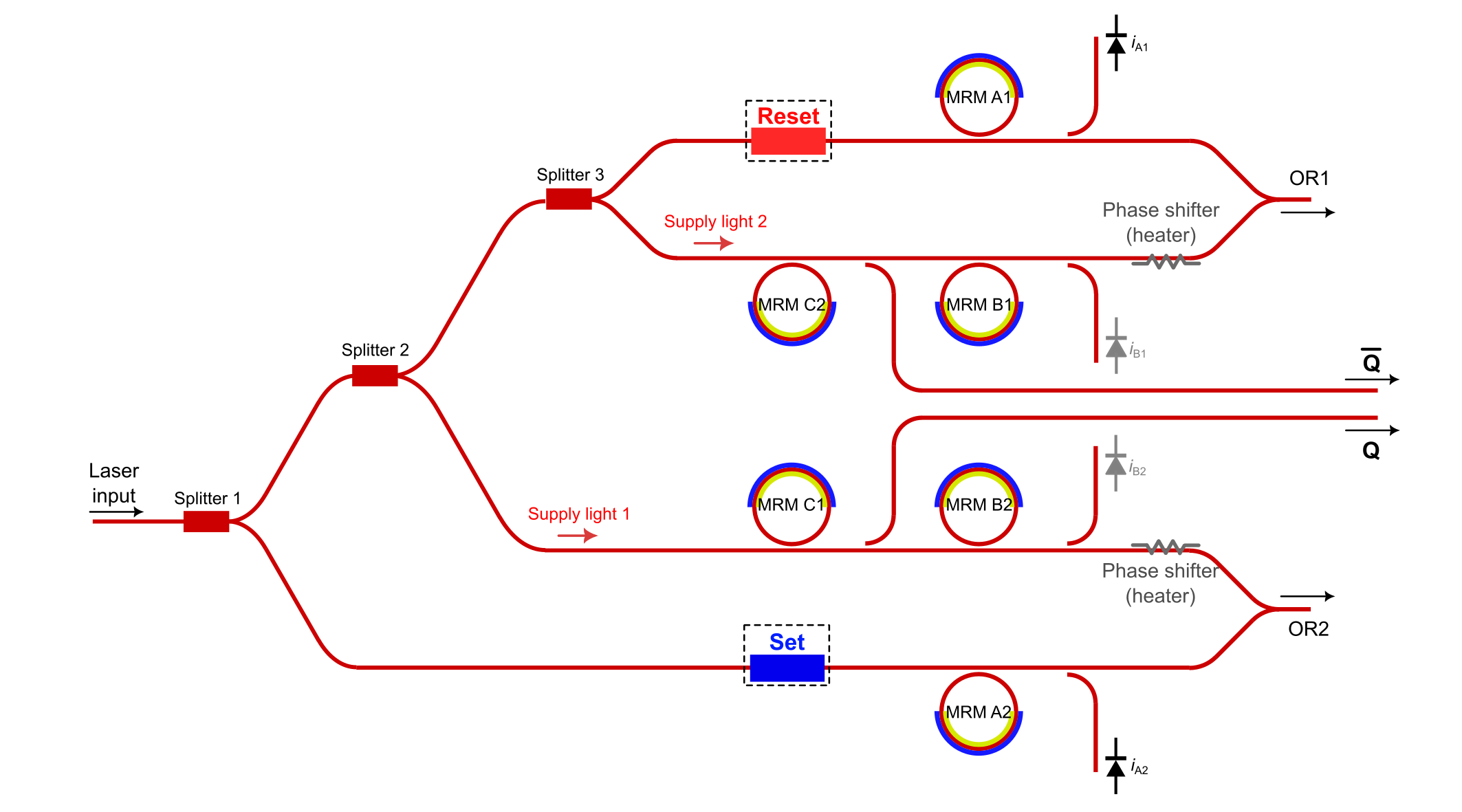}
\caption{Schematic of the SR latch circuit implemented on the simulator mesh shown in Fig. 4a of the main text.}
\label{Fig_S2}
\end{figure}

Similar to the NOR/NAND gate, we use a single optical input to the simulator mesh to implement the SR latch. In Fig. \ref{Fig_S2}, the input light is split into four parts. Splitter 1 generates the set signal, splitter 2 generates supply light 1 to MRM C1, and splitter 3 generates supply 2 to MRM C2 as well as the reset signal. Each NOR gate has two inputs, the first is set/reset, and second comes from the output of the other gate. Note that set and reset states are controlled using two attenuators. Here, the cross-coupled architecture of an SR latch is realized by cascading MRM C1 (C2) and MRM B2 (B1). Essentially, outputs OR1 and OR2 are inverted using MRM C1 and C2, respectively, and each inverted output is coupled directly to one input (MRM B in this case) of the other gate by cascading the modulators. The latch outputs monitored after MRMs C1 and C2 using two couplers. The circuit for each gate is similar to Fig. \ref{Fig_S1}.

\end{document}